\documentclass[11pt]{article}
\usepackage{blois}
\usepackage[font=small,tableposition=top]{caption}

\DeclareCaptionLabelFormat{andtable}{#1~#2  \&  \tablename~\thetable}

\def\be{\begin{equation}}
\def\ee{\end{equation}}
\def\bea{\begin{eqnarray}}
\def\eea{\end{eqnarray}}

\def\ifb{\mbox{~fb$^{-1}$\ }}

\newcommand{\Photo}{\includegraphics[height=35mm]{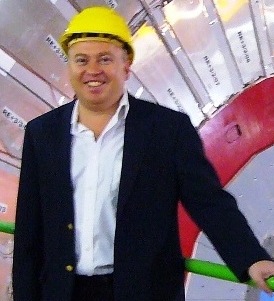}}

\begin{document}
\vspace*{4cm}
\title{LHC: Past, Present, and Future}

\author{ Greg Landsberg }

\address{Brown University, Department of Physics, 182 Hope St,\\
Providence, RI 02912, USA}

\maketitle\abstracts{
In this overview talk, I give highlights of the first three years of the LHC operations at high energy, spanning
heavy-ion physics, standard model measurements, and searches for new particles, which culminated in
the discovery of the Higgs boson by the ATLAS and CMS experiments in 2012. I'll discuss what we found
about the properties of the new particle in 10 months since the discovery and then talk about the future
LHC program and preparations to the 2015 run at $\sqrt{s} \sim 13$ TeV. These proceedings are meant to 
be a snapshot of the LHC results as of May 2013~--- the time of the conference. Many of the results shown
in these proceedings have been since updated (sometimes significantly) just 4 months thereafter, when these 
proceedings were due. Nevertheless, keeping this writeup in sync with the 
results shown in the actual talk has some historical value, as, for one, it tells the reader how short is 
the turnaround time to update the results at the LHC. To help an appreciation of this fact, I briefly summarize 
the main changes between May and September 2013 in the Appendix.
}

\section{Introduction}

The Large Hadron Collider (LHC) is the most powerful particle accelerator built to date. Straddling the French-Swiss border near Geneva, Switzerland, its 27 km ring contains two intense beams of protons ($p$) accelerated to the energy of 3.5--4.0 TeV each and circulating in opposite directions. The beams are kept apart inside an evacuated beam-pipe and are bent in opposite directions by specially designed superconducting magnets. The beams are brought together and collide head-on in four points (\mbox{P}) around the LHC ring, where four large LHC experiments~--- ALICE (P2), ATLAS (P1), CMS (P5), and LHCb (P8)~--- are located, augmented by three smaller special-purpose experiments: LHCf (P1), MoEDAL (P8), and TOTEM (P5). The ATLAS and CMS experiments are general-purpose detectors, capable of exploring all aspects of the LHC program, from heavy-ion collisions and forward physics to Higgs boson physics and direct searches for new particles. The ALICE experiment is dedicated to heavy-ion physics, while LHCb is designed to maximize the LHC potential in B and charm physics. The LHCf and TOTEM experiments measure very forward physics phenomena, with LHCf focusing on measurements of particle production in very forward region and TOTEM~--- on measuring elastic, inelastic, and total $pp$ cross sections. Finally, the MoEDAL experiment is dedicated to searches for magnetic monopoles.

First $pp$ collisions at the LHC took place November 23, 2009, at the beam injection energy of 0.45 TeV ($\sqrt{s} = 0.9$ TeV), and soon thereafter (March 30, 2010) the LHC started to operate at the proton-proton center-of-mass energy $\sqrt{s} = 7$ TeV, which is half of the design energy of the machine. In 2010-2011, an integrated luminosity of $\sim 5\ifb$ to the ATLAS and CMS experiments, and $\sim 1.1\ifb$ to the LHCb experiment was delivered. On April 5, 2012, the machine started its successful $\sqrt{s} = 8$ TeV run, which delivered an integrated luminosity of $\sim 25\ifb$ to each of ATLAS and CMS and $\sim 2.2\ifb$ to LHCb. The peak instantaneous luminosity of $7.7 \times 10^{33}$ cm$^{-2}$s$^{-1}$ was reached by the LHC on November 30, 2012. This is close to the design luminosity of $10^{34}$ cm$^{-2}$s$^{-1}$, albeit at half the design energy and twice the beam crossing time.

The first LHC heavy-ion PbPb collisions took place in 2010 at a center-of-mass energy per pair of colliding nucleons $\sqrt{S_{NN}} = 2.76$~TeV, with an integrated luminosity of $\sim 10~\mu$b$^{-1}$ delivered to each of ALICE, ATLAS, and CMS. Some 15 times more data at the same energy were delivered in 2011. In September 2012 the LHC provided a short, 4-hour long ``pilot" run of $p$Pb collisions at $\sqrt{s_{NN}} = 5.02$ TeV, successfully demonstrating its capability to provide asymmetric collisions. A three-week long $p$Pb run of 2013 delivered data corresponding to an integrated luminosity of $\sim 30$ nb$^{-1}$ to ALICE, ATLAS, and CMS, and $\sim 2$ nb$^{-1}$ to LHCb, which just joined the heavy-ion program, thanks to asymmetric collision conditions. In addition, the LHC also delivered two sets of ``reference'' $pp$ data at $\sqrt{s} = 2.76$ TeV in 2011 and 2013 for comparison with PbPb results at the same per-nucleon collision energy. These data-taking periods comprise the LHC Run 1, which has successfully ended on February 16, 2013, bringing the LHC to the first long shutdown (LS1) necessary for its energy upgrade to the design specifications.

The LHC detectors worked spectacularly well during the LHC Run 1, with more than 95\% of channels operational for ATLAS and CMS after the three-year-long run. The experiments recorded $\sim 95\%$ of data delivered by the LHC and the publications are typically based on 85\%--90\% of the delivered data, which is a remarkable achievement. In addition, both ATLAS and CMS quickly learned how to cope successfully with the large pileup (multiple interactions per beam crossing), which has increased from a negligible level in 2010 to an average of 9 (21) interactions per beam crossing in 2011 (2012). Despite this large pileup, the particle identification in both experiments was tuned to have very little dependence on the number of pileup events, thanks to excellent tracking detectors and high-granularity calorimeters of ATLAS and CMS.

\section{The LHC Legacy}

The LHC has in fact (allegorically speaking!) replaced three machines in one go: the Tevatron (in Higgs boson and beyond-standard-model (BSM) searches, top physics, and precision electroweak measurements), Belle (precision B-physics), and RHIC (heavy-ion physics). The LHC experiments are very successful in all three areas, which would not have been possible without theoretical and phenomenological breakthroughs of the past decade: higher-order calculations, modern Monte Carlo generators, reduced parton distribution function uncertainties, etc. This introductory talk will highlight just a few of the most exciting results of the past three years, with many more covered in the dedicated plenary and parallel talks in these proceedings.

Here are some milestones that the LHC has achieved over its rather short operational history:
\begin{itemize}
\item September 10, 2008: the first beams were injected and circulated in the LHC machine; the beam lifetime was measured to be within the design specifications.
\item November 28, 2009: the first LHC paper \cite{ALICE1} by the ALICE Collaboration on the measurement of charged-particle pseudorapidity density at $\sqrt{s} = 0.9$ TeV has been submitted to Eur. Phys. J. C.
\item September 29, 2010: the CMS ``ridge" paper \cite{CMS-ridge} that observed a new and unexpected phenomenon of long-range near-side two-particle correlations in high-multiplicity $pp$ collisions at $\sqrt{s} = 0.9$, 2.36, and 7 TeV, has been submitted to JHEP.
\item December 21, 2011: the ATLAS paper \cite{ATLAS-chib} claiming an observation of the first new particle at the LHC, an excited $\chi_b(3P)$ baryon, has been submitted to Phys. Rev. Lett.
\item July 4, 2012: a new boson discovery has been announced at a special CERN seminar by the ATLAS and CMS Collaborations; papers have been submitted to Phys. Lett. B journal on July 31, 2012~\cite{Higgs-discovery}.
\item November 12, 2012: the LHCb paper~\cite{Bsmm} claiming the first evidence for a $B^0_s(\mu\mu)$ decay consistent with the standard model (SM) expectations has been submitted to Phys. Rev. Lett.
\end{itemize}
We are looking forward to expanding this set of milestones in 2013 and beyond.

\section{Flavor Physics Results}

The flavor physics program at the LHC is lead by the dedicated LHCb experiment, while the ATLAS and CMS experiments are significant contributors in selected topics. Among the flavor physics highlights of the LHC Run 1 are:
\begin{itemize}
\item Observation of new heavy excited beauty bound states: $\chi_b(3P)$ $b\bar b$ meson~\cite{ATLAS-chib} and $\Xi^*_b$ baryon~\cite{CMS-Xib};
\item Measurement of $\Upsilon(nS)$ and $\psi$ polarization~\cite{CMS-polar}; 
\item First evidence for the $B^0_s \to \mu\mu$ decay~\cite{Bsmm};
\item First observation of direct CP violation in $B^0_s$ decays~\cite{LHCb-CP};
\item Strong constraints on new physics in the bottom and charm sectors via precision measurement of a number of rare decays.
\end{itemize}

In particular, the three-decade-long quest to find a deviation from the SM predictions in the rare flavor-changing-neutral-current $B^0_s(\mu\mu)$ decay seems to be coming to an end with the LHCb 3.5$\sigma$ evidence for observing this decay and the measurement of its branching fraction $B(B^0_s \to \mu\mu) = 3.2^{+1.5}_{-1.2} \times 10^{-9}$. The analysis also set a stringent limit on the $B^0(\mu\mu)$ decay $B(B^0 \to \mu\mu) < 9.4 \times 10^{-10}$ at a 95\% confidence level (CL)~\cite{Bsmm}. The $B^0_s(\mu\mu)$ decay evidence still awaits for a confirmation from ATLAS or CMS.

There are also important new results on CP-violation in the heavy-flavor sector:
\begin{itemize}
\item Most precise determination of $B^0_s$ oscillation parameters using the $\pi^\pm D_s^\mp$ mode~\cite{LHCb-Gs}: $\Delta m_s = 17.768 \pm 0.023~\mbox{(stat.)} \pm 0.006~\mbox{(syst.)}$ ps$^{-1}$; 
\item First observation of direct CP violation in the $B^0_s \to K\pi$ decays, $A_{CP}(B^0_s \to K\pi) = \frac{\Gamma(\overline{B}^0_s \to K^+\pi^-) - \Gamma(B^0_s \to K^-\pi^+)}{\Gamma(\overline{B}^0_s \to K^+\pi^-) + \Gamma(B^0_s \to K^-\pi^+)} = 0.27 \pm 0.04~\mbox{(stat.)} \pm 0.01~\mbox{(syst.)}$~\cite{LHCb-CP};
\item Despite earlier hopes, no evidence was found for large CP violation in $B^0_s \to J/\psi\phi$ decays: $\phi_s = 0.12 \pm 0.25~\mbox{(stat.)} \pm 0.11~\mbox{(syst.)}$, ATLAS; $\phi_s = 0.07 \pm 0.09~\mbox{(stat.)} \pm 0.01~\mbox{(syst.)}$, LHCb~\cite{noCPVJpsiphi};
\item Lack of confirmation by new LHCb analyses~\cite{noCPVcharm} of the previously reported evidence for direct CP violation in the $D^0 \to \pi^+\pi^-,\; K^+K^-$ decays ($\Delta A_{\rm CP} = -0.678 \pm 0.147$, 4.6$\sigma$ away from 0~\cite{HFAG2012}); the new world average value is $\Delta A_{\rm CP} = -0.329 \pm 0.121$, i.e. $<3\sigma$ away from 0~\cite{HFAG2013}.
\end{itemize}

\section{Heavy-Ion Results}

Very successful PbPb (2010, 2011) and $p$Pb (2013) runs brought a wealth of data and allowed ALICE, ATLAS, and CMS to produce unprecedented and very exciting new results:
\begin{itemize}
\item Detailed studies of jet quenching in PbPb collisions~\cite{JetQuenching} and dijet production in $p$Pb collisions~\cite{CMSdijetspPb};
\item Elliptic flow and multiparticle correlations including studies of the ``ridge" in $pp$, $p$Pb, and PbPb collisions~\cite{ridge};
\item $\Upsilon(2S)$ and $\Upsilon(3S)$ ``melting" in PbPb collisions~\cite{CMSUpsilon};
\item Number of other unique PbPb measurements:
\begin{itemize}
\item W and Z production~\cite{WZPbPb};
\item Jet-photon correlations~\cite{jgPbPb};
\item Charm suppression~\cite{ALICE-charm};
\item Nuclear modification factor for b-tagged jets~\cite{CMSHIbtag}.
\end{itemize}
\end{itemize}
The new $p$Pb data taken earlier this year are being analyzed very rapidly and already showed a lot of unexpected features that have a potential to shift the very paradigm in heavy-ion physics. The LHCb experiment is now joining the program with the $J/\psi$ suppression measurement in $p$Pb collisions at forward rapidities~\cite{LHCBHI}.

\section{TOTEM and LHCf Results}

The TOTEM Collaboration has recently completed an important program of luminosity independent measurements of elastic, inelastic, and total $pp$ cross sections at $\sqrt{s} = 7$ and 8 TeV~\cite{TOTEM}, summarized in Table~\ref{table:pp}.

\begin{table}[tbh]
\caption{Luminosity independent $pp$ cross section measurements from TOTEM experiment}
\label{table:pp}
\vspace{-0.4cm}
\begin{center}
\begin{tabular}{|l|c|c|c|}
\hline
$\sqrt{s}$ & Total cross section & Elastic cross section  & Inelastic cross section \\
\hline
7 TeV & $98.0 \pm 2.5$ mb & $25.4 \pm 1.1$ mb & $73.2 \pm 1.3$ mb\\ 
\hline
8 TeV & $101.7 \pm 2.9$ mb & $27.1 \pm 1.4$ mb & $74.7 \pm 1.7$ mb\\
\hline
\end{tabular}
\end{center}
\vspace{-0.2cm}
\end{table}

The LHCf collaboration made several measurements of particle production in a very forward region ($8 < y < 15$). These measurements provide an important input for tuning Monte Carlo generators used for description of showers induced by ultrahigh energy cosmic rays. The new measurement from LHCf~\cite{LHCf} prefers {\sc EPOS 1.99} model~\cite{EPOS} for description of the transverse momentum spectrum of $\pi^0$ mesons produced at these rapidities.

\section{The Higgs Boson Story}

After the fireworks of July 4, 2012 and the announcement of the discovery of a new boson~\cite{Higgs-discovery}, both ATLAS and CMS set off on a long path of measuring various properties of the new particle and determining if this is the long-sought SM Higgs boson. The amount of work commenced over just ten months since the discovery is quite remarkable. For most of the channels, the full Run 1 statistics have been analyzed, which amounts to 2.5 times the discovery sample. Here are the most important findings of the past year:
\begin{itemize}
\item The existence of new particle has been established beyond any doubts~\cite{ATLAS-Higgs,CMS-Higgs} (see Fig.~\ref{fig:ATLASsig} and Table~\ref{table:CMSsig}); 
\item It is a $J^{PC} = 0^{++}$ boson responsible for EWSB, as evident from its relative couplings to $W/Z$ bosons vs. photons~\cite{ATLAS-Higgs,CMS-Higgs};
\item Its properties are consistent with those of the SM Higgs boson within (sizable) uncertainties~\cite{ATLAS-Higgs,CMS-Higgs}; 
\item There is mounting evidence~\cite{CMS-Higgs,Tevatron} that it couples to at least third-generation, down-type fermions.
\end{itemize}

The Higgs boson mass has been measured to a remarkable precision: 0.43\% in ATLAS~\cite{ATLAS-Higgs} and 0.34\% in CMS~\cite{CMS-Higgs}. Its mass is already know to a better relative precision than the mass of the top (or any other) quark! Similarly, the production cross section relative to its SM prediction, $\mu = \sigma/\sigma_{\rm SM}$ has been measured to $\sim 15\%$ precision by each experiment, with the central values of $\mu$ consistent with unity within 1.5$\sigma$~\cite{ATLAS-Higgs,CMS-Higgs}. The data strongly prefer the spin-parity for a new particle to be consistent with that of the vacuum ($J^{PC} = 0^{++}$), which is also the value predicted for the SM Higgs boson. Each experiment ruled out the pseudoscalar~\cite{CMS-Higgs,ATLAS-PS} and tensor~\cite{CMS-Higgs,ATLAS-2} hypotheses at 2--3$\sigma$ level. The vector and pseudovector hypotheses are ruled out by the observation of the new boson in the $\gamma\gamma$ decay mode, as a consequence of the Landau-Yang theorem~\cite{LY}. These results are summarized in Table~\ref{table:Higgs} and indicate significant progress in understanding of the properties of the new particle since its discovery. So far, it does look more and more like the SM Higgs boson! Measurements of the $\mu$ values for the new boson in the individual bosonic and fermionic channels, summarized in Fig.~\ref{fig:couplings}, lead to the same conclusion.
\medskip

\noindent
\begin{minipage}{\textwidth}
\begin{minipage}[t]{0.49\textwidth}
\centering
    \includegraphics[width=1.0\linewidth]{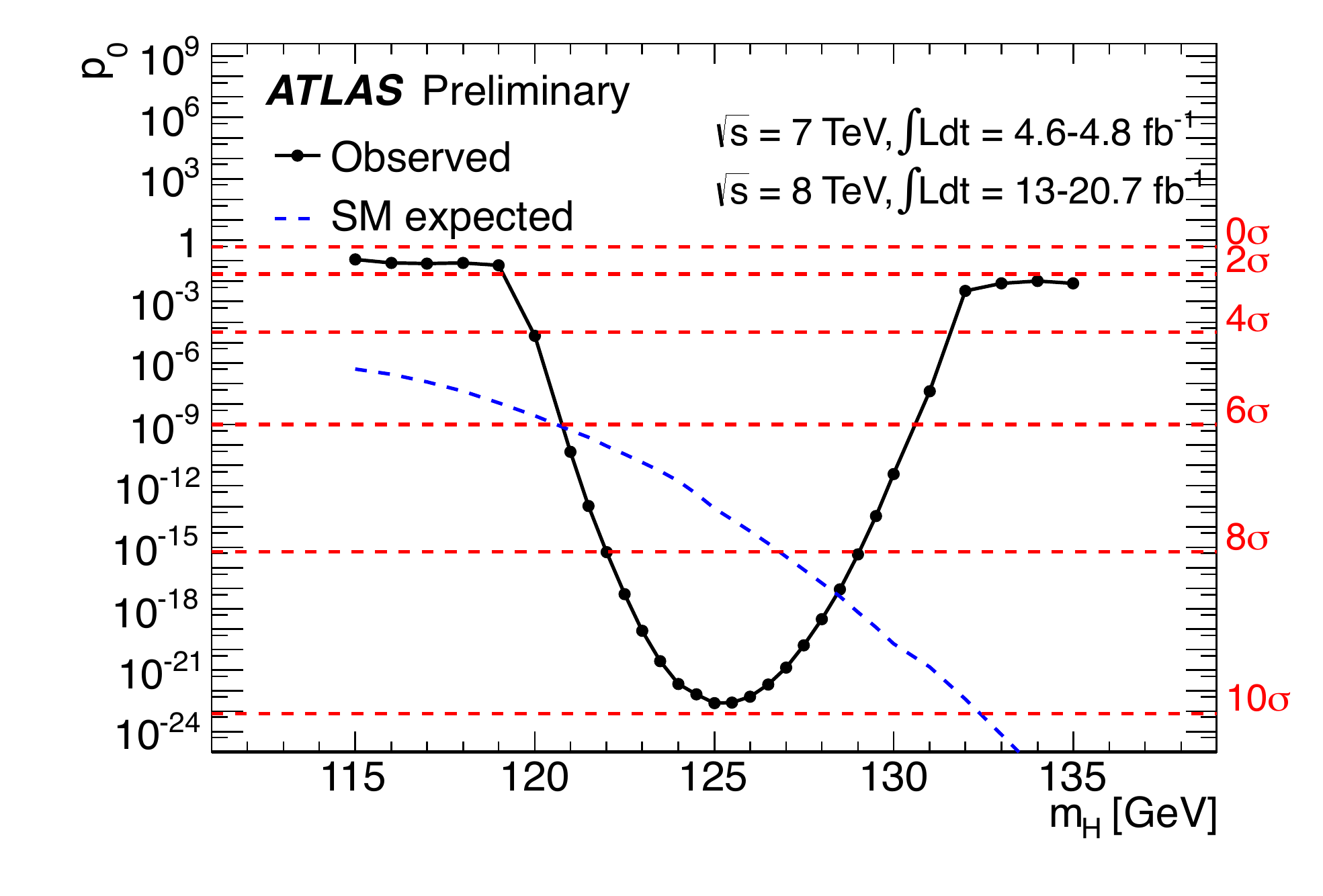}
    \vspace*{-1.0cm}
    \captionof{figure}{Combined Higgs boson observation significance in ATLAS~\protect\cite{ATLAS-Higgs}.}
    \label{fig:ATLASsig}
\end{minipage}
\hfill
\begin{minipage}[b]{0.49\textwidth}
\centering
\vspace*{-1.8cm}
    \captionof{table}{Expected and observed Higgs boson observation significances in various channels in CMS~\protect\cite{CMS-Higgs}.}
    \label{table:CMSsig}
    \vspace*{-0.2cm}
    \begin{tabular}{|l|c|c|}
    \hline
    Channel & Expected, $\sigma$ & Observed, $\sigma$ \\
    \hline
    $ZZ$ & 7.1 & 6.7 \\
    $\gamma\gamma$ & 3.9 & 3.2 \\
    $WW$ & 5.3 & 3.9 \\
    $b\bar b$ & 2.2 & 2.0 \\
    $\tau\tau$ & 2.6 & 2.8 \\
    $b\bar b + \tau\tau$ & 3.4 & 3.4 \\
    \hline
    \end{tabular}
  \vspace{0.6cm}
    \end{minipage}
    \bigskip
  \end{minipage}
  
The measured value of the Higgs boson mass is quite interesting in itself. Indeed, the new particle is light enough to be a minimal supersymmetric standard model (MSSM) Higgs boson; yet it is too heavy to obviously prefer MSSM over SM~\cite{Sven}. Vacuum stability arguments seem to point to the scale $\sim 10^{8}$ TeV below which some new physics must exist. Curiously, a similar scale is suggested by the observed neutrino mass hierarchy, assuming a see-saw mechanism for the neutrino masses. On the other hand, a metastable vacuum may survive all the way to the Planck scale~\cite{vacuum}, as seen in Fig.~\ref{fig:vacuum} (left). The simultaneous measurement of the Higgs boson and top-quark masses allowed for the first time to infer the properties of the very vacuum we live in! We find ourselves in a ``just-so" situation: the vacuum is at the verge of being either stable or metastable. A sub-percent change of $\sim 1$ GeV in either the top-quark or the Higgs boson mass is all it takes to tip the scales! Perhaps Nature is trying to tell us something here? It is therefore very important to improve on the precision of top-quark mass measurements, including various complementary methods and reduction of theoretical uncertainties. Tevatron is still leading with the new combined top-quark mass result, but the LHC is catching up quickly~\cite{top-mass}, as shown in Fig.~\ref{fig:vacuum} (right).

If the new boson is the SM Higgs boson, we have to elucidate the very mechanism of electroweak symmetry breaking, i.e., explain the origin of the quartic term in the Higgs potential. If the new boson is an MSSM Higgs boson, we need to find other supersymmetric particles and other Higgs bosons. In a sense, a $\sim 126$ GeV Higgs boson is maximally challenging and rich experimentally, but it also inflicts ``maximum pain" theoretically, as it is not so easy to accommodate.

\begin{table}[htb]
\caption{Properties of the Higgs boson from global analysis of various production and decay channels.}
\label{table:Higgs}
\vspace{-0.6cm}
\begin{center}
\begin{tabular}{|l|c|c|}
\hline
Property & ATLAS & CMS \\
\hline
Mass & $125.5 \pm 0.2^{+0.5}_{-0.6}$~GeV & $125.7 \pm 0.3 \pm 0.3$~GeV \\
$\mu$ & $1.30 \pm 0.20$ @ 125.5 GeV & $0.80 \pm 0.14$ @ 125.7 GeV \\
\hline
CL$_S:\; J^{PC} = 0^{-+}$ vs. $0^{++}$ hypothesis & 0.022 & 0.016 \\
CL$_S:\; J^{PC} = 2^{++}$ vs. $0^{++}$ hypothesis & $<4 \times 10^{-4}$ & $<0.015$ \\
\hline
\end{tabular}
\end{center}
\vspace*{-0.4cm}
\end{table} 

\begin{figure}[hbt]
\centering
    \includegraphics[width=0.49\linewidth]{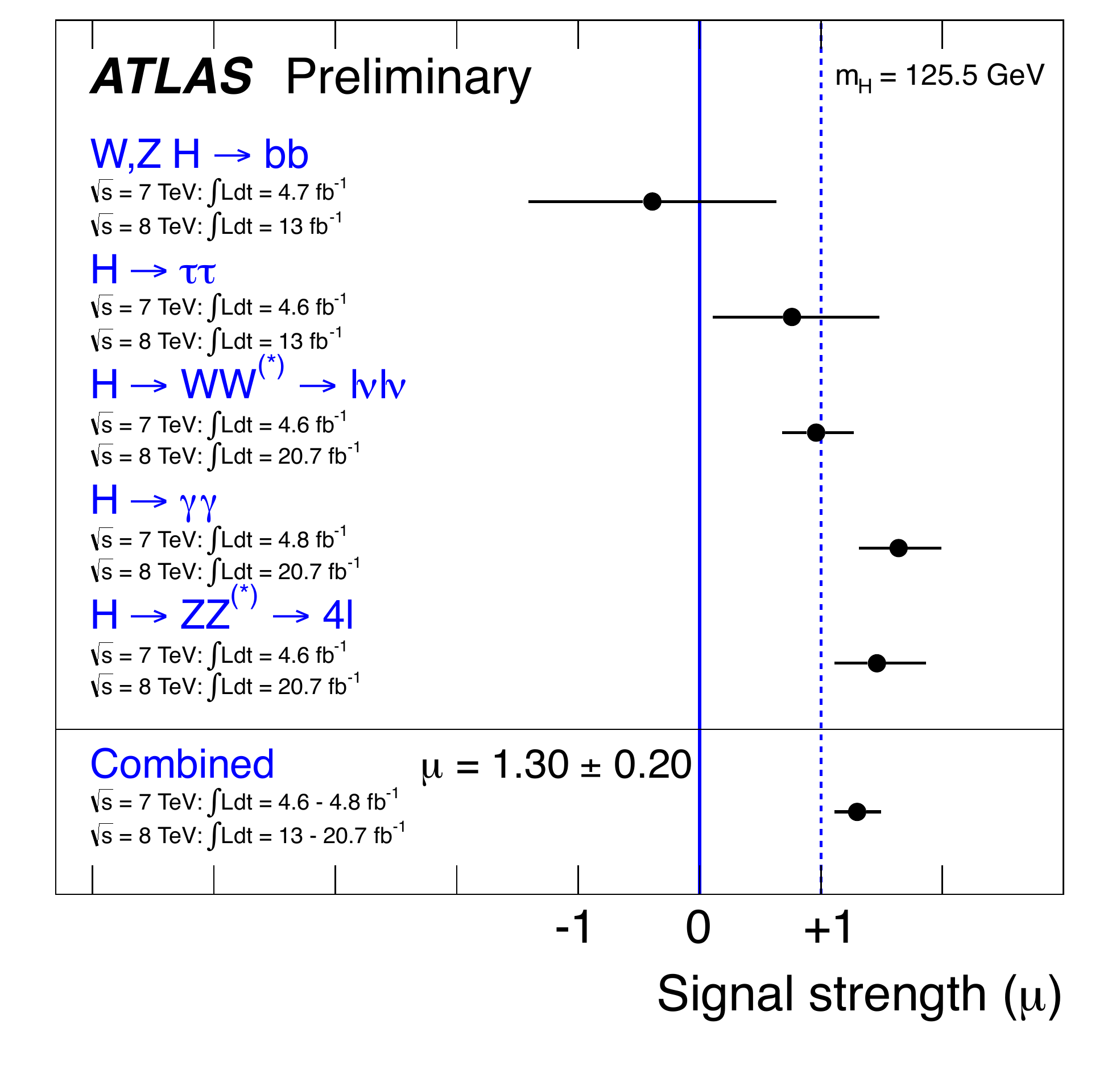}\hfill
    \includegraphics[width=0.44\linewidth]{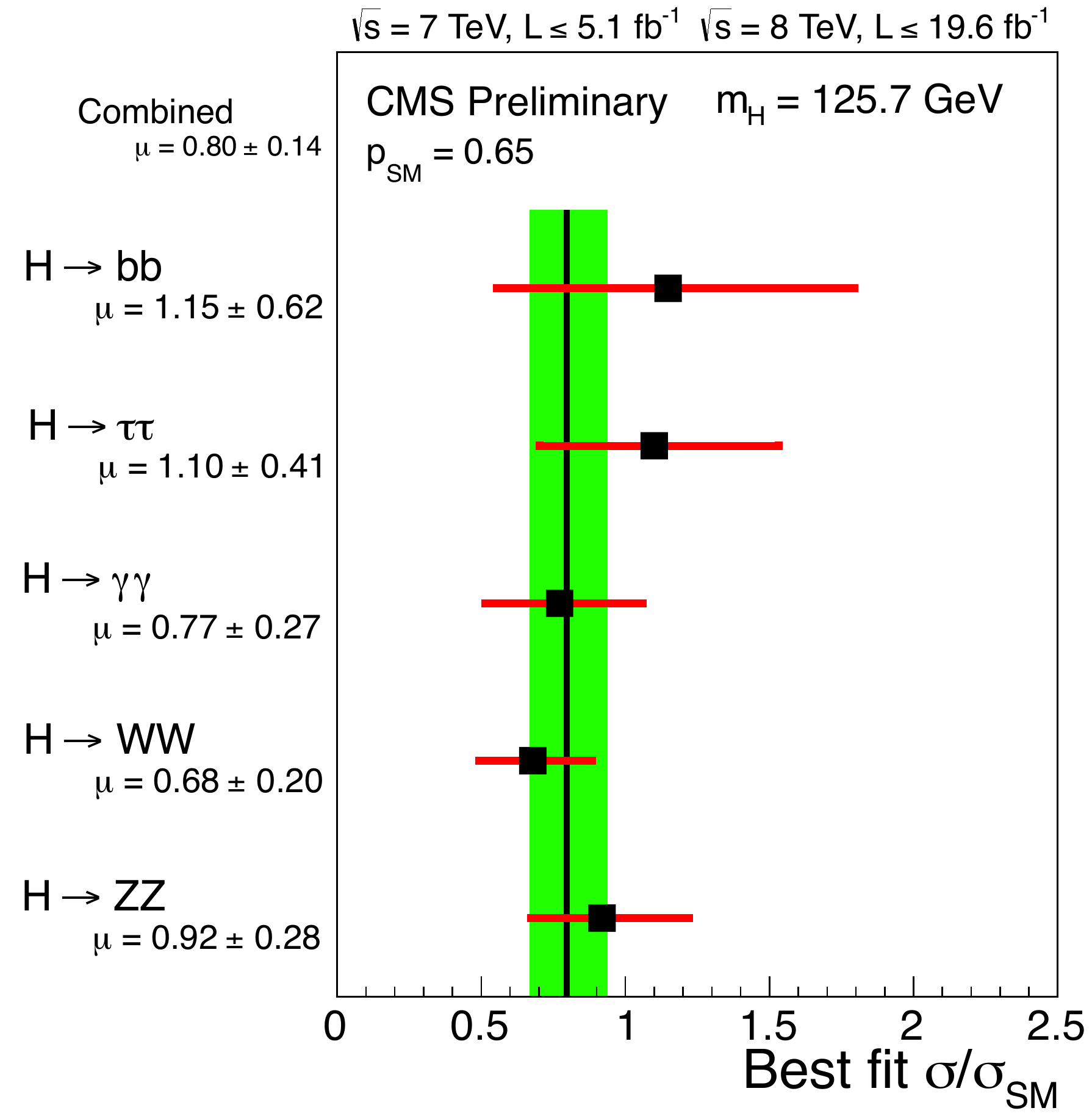}
    \vspace*{-0.4cm}
    \captionof{figure}{Measurements of the Higgs boson $\mu$ values in the individual channels by the (left) ATLAS~\protect\cite{ATLAS-Higgs} and (right) CMS~\protect\cite{CMS-Higgs} experiments.}
\label{fig:couplings}
\vspace*{-0.6cm}
\end{figure}

\begin{figure}[hbt]
\centering
    \includegraphics[width=0.55\linewidth]{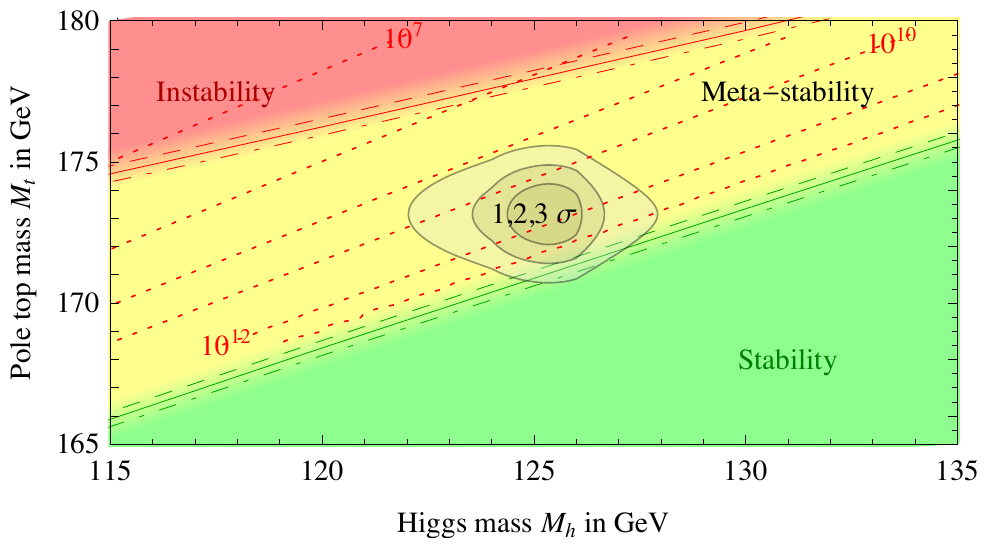}\hfill
    \includegraphics[width=0.40\linewidth]{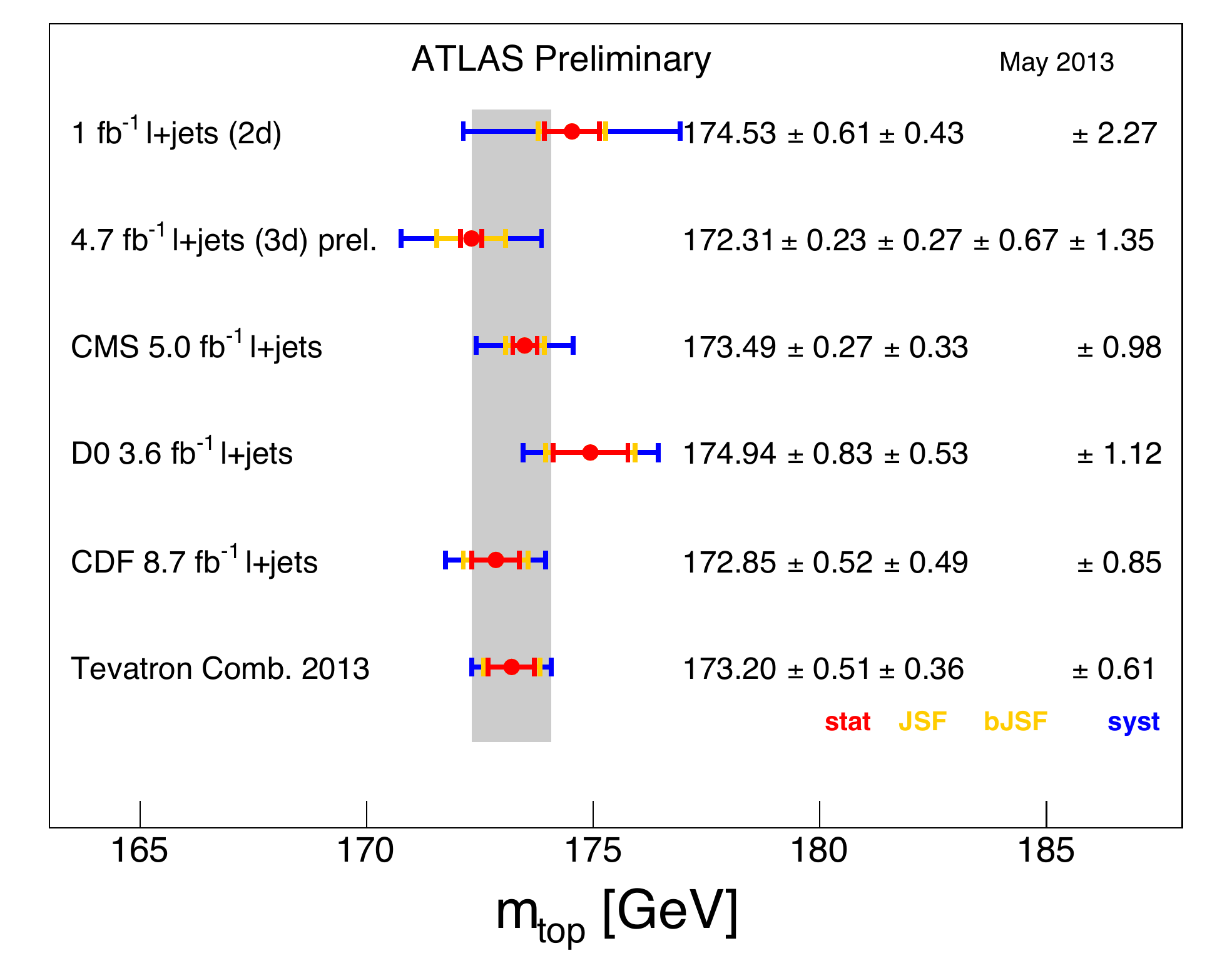}
    \vspace*{-0.4cm}
    \captionof{figure}{(Left) Constraints on vacuum stability as a function of the Higgs boson and top-quark masses. (From Ref.~\protect\cite{vacuum}.) (Right) Measurements~\protect\cite{top-mass} of the top-quark mass at the Tevatron and the LHC.}
\label{fig:vacuum}
\vspace*{-0.4cm}
\end{figure}

\section{Searches for New Physics}

While the Higgs boson has been successfully discovered last year, many searches for BSM physics so far have revealed nothing new. This is not due to a lack of them; in fact, the LHC collaborations, led by ATLAS and CMS, have searched for all kinds of new phenomena, from supersymmetry to black holes, from technicolor to vector-like quarks. Stringent limits on many models have been set, and some of them, e.g., technicolor or fourth-generation quarks, are all but ruled out by these searches. The big picture of ATLAS and CMS searches for various phenomena can be found in Refs.~\cite{ATLAS-searches,CMS-searches}.

Of particular interest are extensive searches for supersymmetry (SUSY), a theory that potentially offers an elegant solution to the hierarchy problem of the SM, provides an excellent dark matter (DM) candidate, and unification of gauge couplings. Generic searches for squarks and gluinos traditionally cast in a constrained version of MSSM, which reduces the number of free SUSY parameters to five by making various unification assumptions, have excluded~\cite{ATLAS-cMSSM} squarks (gluinos) to the masses of about 2.0 TeV (1.1 TeV) (see Fig.~\ref{fig:cMSSM}, left), which is already higher than can be comfortably accommodated by the model. Does it mean that our hopes to find SUSY are all but vanished? In fact, it doesn't!

What we have excluded so far is the simplest, overconstrained SUSY model, which has many ad hoc assumptions to reduce the number of free parameters. It was a convenient framework for casting sensitivity of different searches in the past, but it clearly overgrew its usefulness. Simple change in just one of the rather arbitrary assumptions~---- the mass degeneracy of squarks~--- reduces the limits on the lightest squark mass by more than a factor of two~\cite{alphaT}, see Fig.~\ref{fig:cMSSM}, right!

\begin{figure}[hbt]
\centering
    \includegraphics[width=0.54\linewidth]{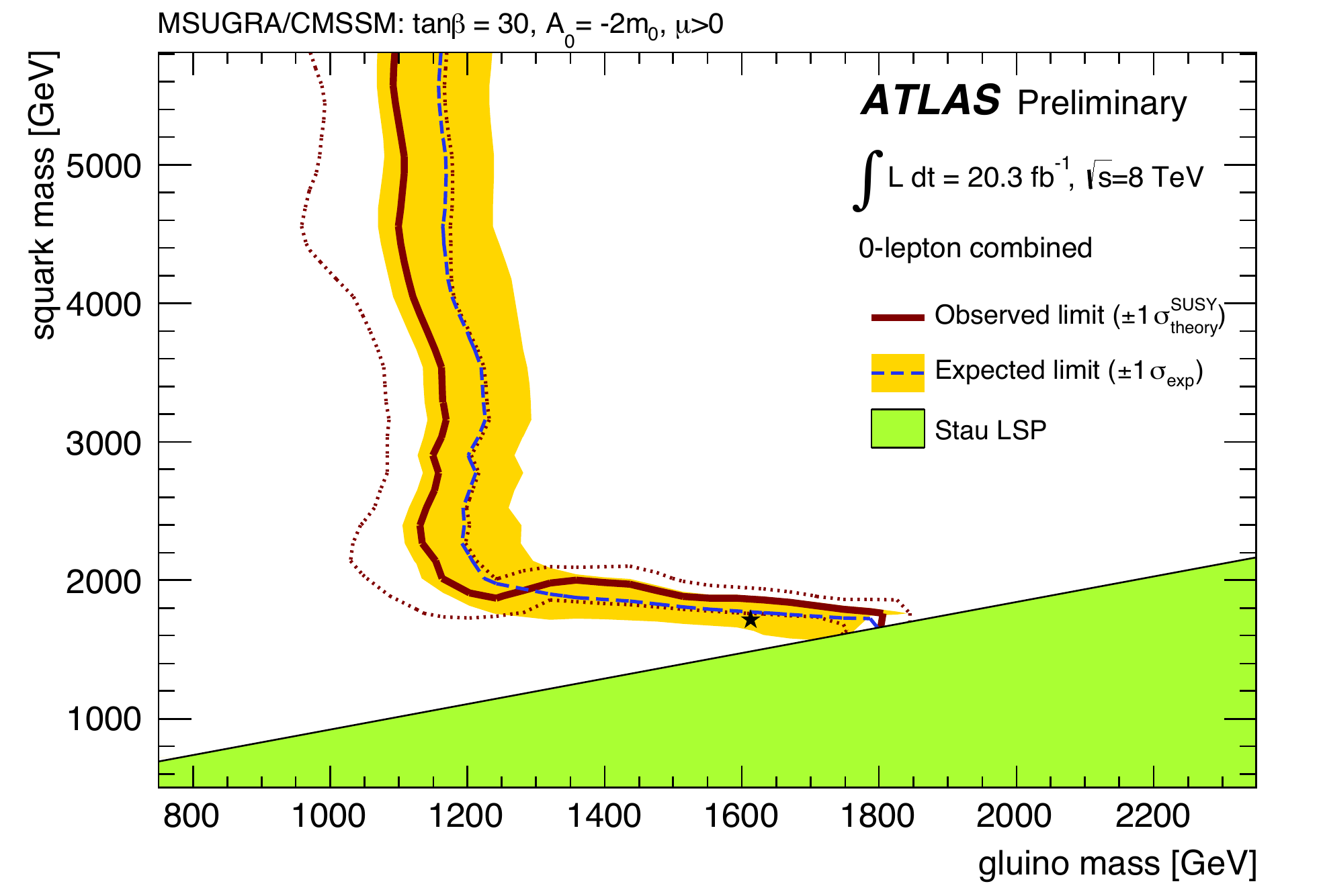}\hfill
    \includegraphics[width=0.45\linewidth]{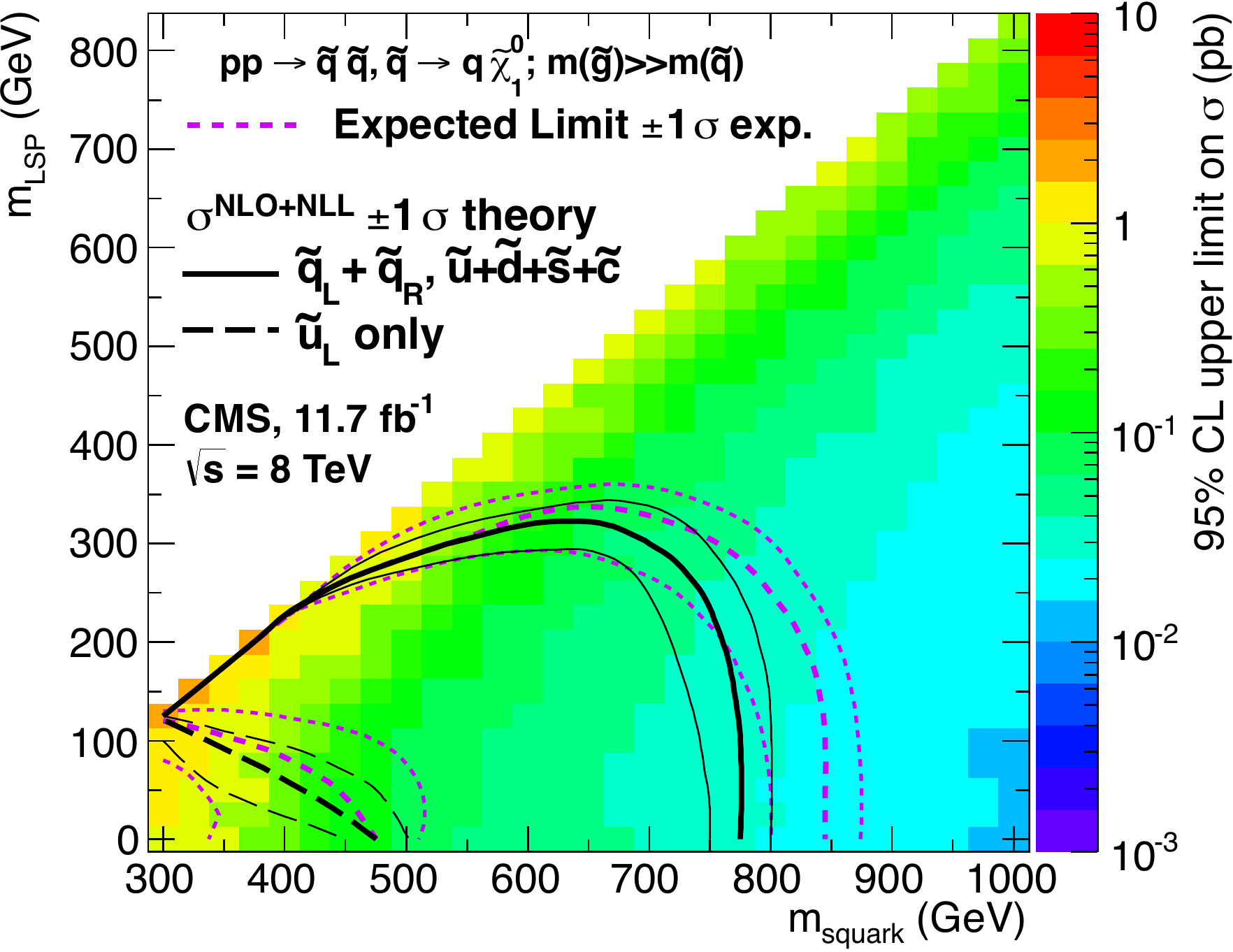}
    \vspace*{-0.4cm}
    \captionof{figure}{Supersymmetry or supercemetery? (Left) Constraints on the gluino vs. squark mass in constrained MSSM framework from ATLAS~\protect\cite{ATLAS-cMSSM}. (Right) Constraints on squark vs. neutralino ($m_{\rm LSP}$) mass for degenerate (solid black line) and non-degenerate (dashed black line) squarks from CMS~\protect\cite{alphaT}.}
\label{fig:cMSSM}
\end{figure}

Consequently, since mid-2011, ATLAS and CMS largely moved to a different way of casting their searches for SUSY~--- via the simplified model space (SMS)~\cite{SMS}, where the properties of the SUSY decays are typically given by the masses and decay branching fractions of just a handful of particles, with all other SUSY particles assumed to be heavy.

Given the observation of the Higgs boson last year, the big question that the LHC is posed to answer is whether the solution to the hierarchy problem is ``natural," i.e., the one that does not require large amount of fine tuning. Natural solutions tend to predict relatively light spectrum of new particles, whereas unnatural ones could easily have new physics mass scale beyond the reach of the LHC. A well-studied example of a natural model is supersymmetry, where naturalness arguments dictate (see, e.g., Ref.~\cite{natural}) that three types of superpartners have to be relatively light: the Higgsinos must be below $\sim 0.5$ TeV, at least one of the chiral states of the bottom squark (sbottom) and both states of the top squark (stop) must be below $\sim 1$ TeV, and gluinos must be below $\sim 1.5$ TeV. All other superpartners can be decoupled, i.e., significantly heavier than 1 TeV. Of course, the exact limits depend on the amount of fine tuning one is willing to allow for, and one particular type of particle can be a bit more massive that these rough limits, but the bottom line is that if all three of them are heavier than these limits, SUSY must be fairly fine-tuned.

Therefore, the focus of the most recent Run 1 SUSY searches at the LHC were searches for these three types of particles. One could either look for stop, sbottom, and Higgsino pair-produced directly, or via decays of pair-produced gluinos. The former processes typically have lower cross section, while the latter one is limited by the energy reach of the machine, which is $\sim 1.3$ TeV on the gluino mass with the current statistics at 8 TeV. Therefore, looking for direct and gluino-mediated production offer two complementary approaches. Figure~\ref{fig:SUSY} illustrates the current limits (coming from a number of different searches) on direct stop pair production from the ATLAS experiment and on gluino-mediated sbottom production from the CMS experiment. While large portions of natural SUSY space have been already ruled out by these searches, there is still a significant uncharted territory that will require the 13 TeV LHC running to explore.

\begin{figure}[hbt]
\vspace*{-0.1cm}
\centering
    \includegraphics[width=0.50\linewidth]{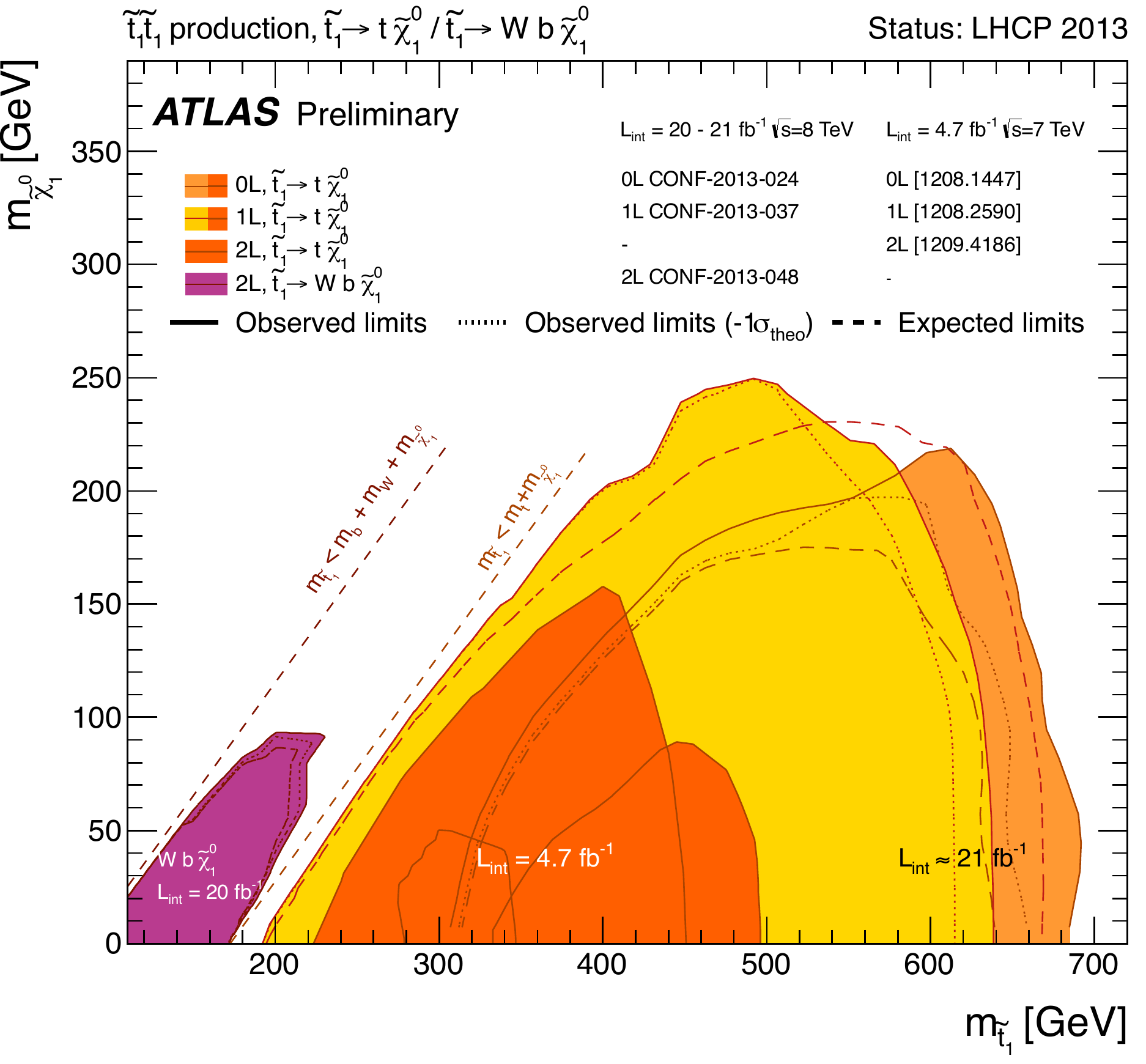}\hfill
    \includegraphics[width=0.48\linewidth]{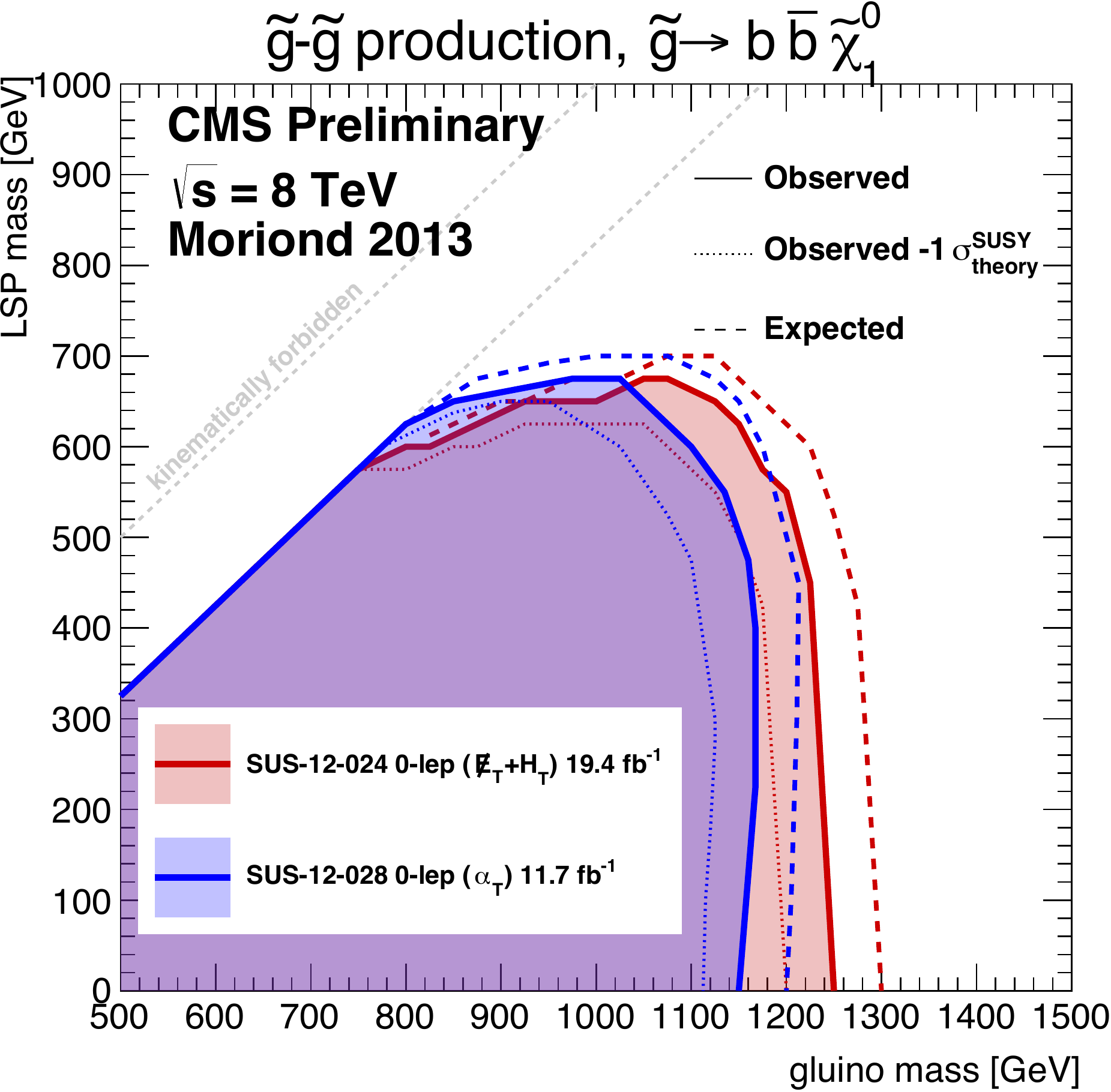}
    \vspace*{-0.4cm}
    \captionof{figure}{(Left) Constraints on the stop vs. neutralino mass from ATLAS~\protect\cite{ATLAS-searches}. (Right) Constraints on gluino-mediated sbottom production from CMS~\protect\cite{CMS-searches}.}
\vspace*{-0.3cm}
\label{fig:SUSY}
\end{figure}

Another interesting way to look for SUSY (or other BSM physics) at the LHC that has a direct connection to astroparticle physics is to search for DM in a process similar to that explored by the direct detection experiments (DDE), which measure nuclear recoil from a DM particle scattering off a nucleus within the detector. If the DM-quark interactions are spin independent, the DDE benefit from the coherent scattering off the entire nucleus, which enhances their sensitivity. For spin dependent interactions, the coherence is lost and therefore the sensitivity is significantly weaker than in the former case. The DDE are typically sensitive to the DM particle masses between $\sim 10$ GeV and a few TeV; light DM detection remains hard, as the nuclear recoil becomes less and less pronounced for DM mass below the nucleus mass, so for a very light DM particle the recoil energy drops below the sensitivity of the present detectors.

The same process that is used to detect DM particles ($\chi$) via nuclear recoil, $\chi q \to \chi q$, can be turned around to look for direct production of pairs of DM particles at the LHC via $q\bar q \to \chi\bar\chi$. This process can be parameterized via an effective field theory four-point operator with an unknown scale (usually denoted either $\Lambda$ or $M^*$ in the literature) related to the interaction mediator mass and its couplings to quarks and DM particles, in a similar same way Fermi's four-point description of weak decays can be related to the W boson mass and its couplings. While the effective theory approach technically applies only to the case of heavy mediator with the mass significantly higher than the mass of the DM particle, it could also be generalized to the case of a light mediator (e.g., the $Z$ boson or the Higgs boson) with certain caveats~\cite{DMcolliders}. 

However, the $q\bar q \to \chi\bar\chi$ process corresponds to an invisible final state, and therefore can not be straightforwardly  detected at colliders. The rescue comes from the initial-state radiation of gluons or photons, which turns this process in a production of a single photon or a jet, recoiling against the DM particles. This is a spectacular signature at the LHC, with an apparent momentum non-conservation due to an emission of DM particles that escape detection. The main irreducible background to this process is a production of a $Z$ boson in association with either a jet or a photon, with the $Z$ boson decaying into neutrinos. Other backgrounds come from QCD production of dijets or direct photons with one of the jets being mismeasured and from $W$ boson production in association with a jet or a photon, with a lepton from the $W$-boson decay being lost. Unlike the DDE, collider searches are nearly equally sensitive to light and heavy DM particles; their sensitivity drops only above the kinematic limit where the DM particle mass becomes so heavy that their pair production at the LHC is suppressed. 

Both ATLAS and CMS have searched for DM production in the monojet and monophoton channels and set stringent limits on the cross section of DM-quark interaction. These searches offer unique sensitivity to spin dependent interactions and for the case of light DM particles, as can be seen from Fig.~\ref{fig:DM}. They can be also used to probe the DM interaction with gluons, which is not accessible by the DDE. These results are particularly interesting in the light of a slight excess consistent with light DM seen by several DDE, most recently by CDMS~\cite{CDMS}.

The same analyses in the monojet and monophoton final states have been also used to set stringent limits on models with large extra dimensions in space and TeV-scale gravity, where the signature comes from the production of a graviton recoiling against a jet or a photon.

\begin{figure}[hbt]
\vspace*{-0.3cm}
\centering
    \includegraphics[width=0.53\linewidth]{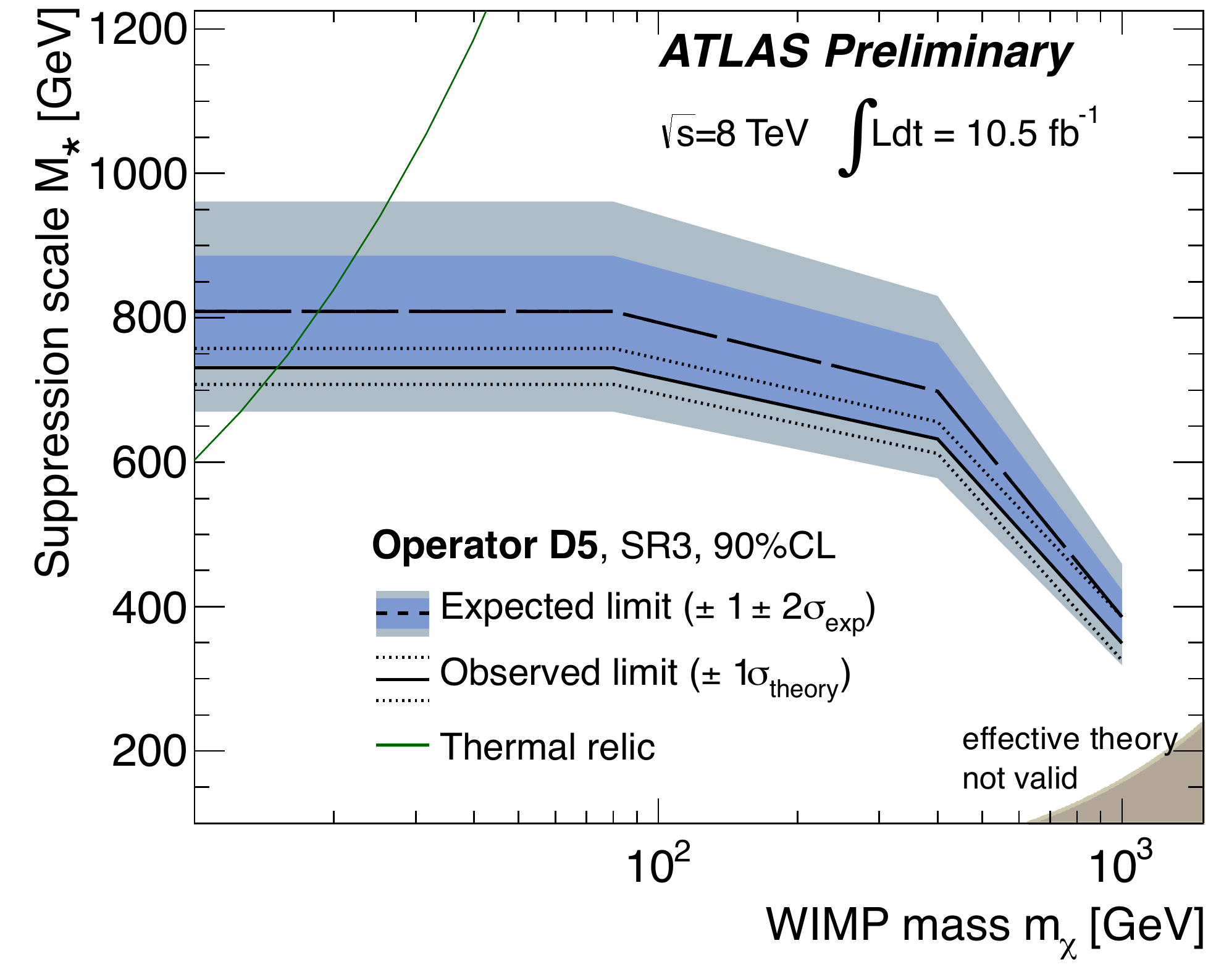}\hfill
    \includegraphics[width=0.46\linewidth]{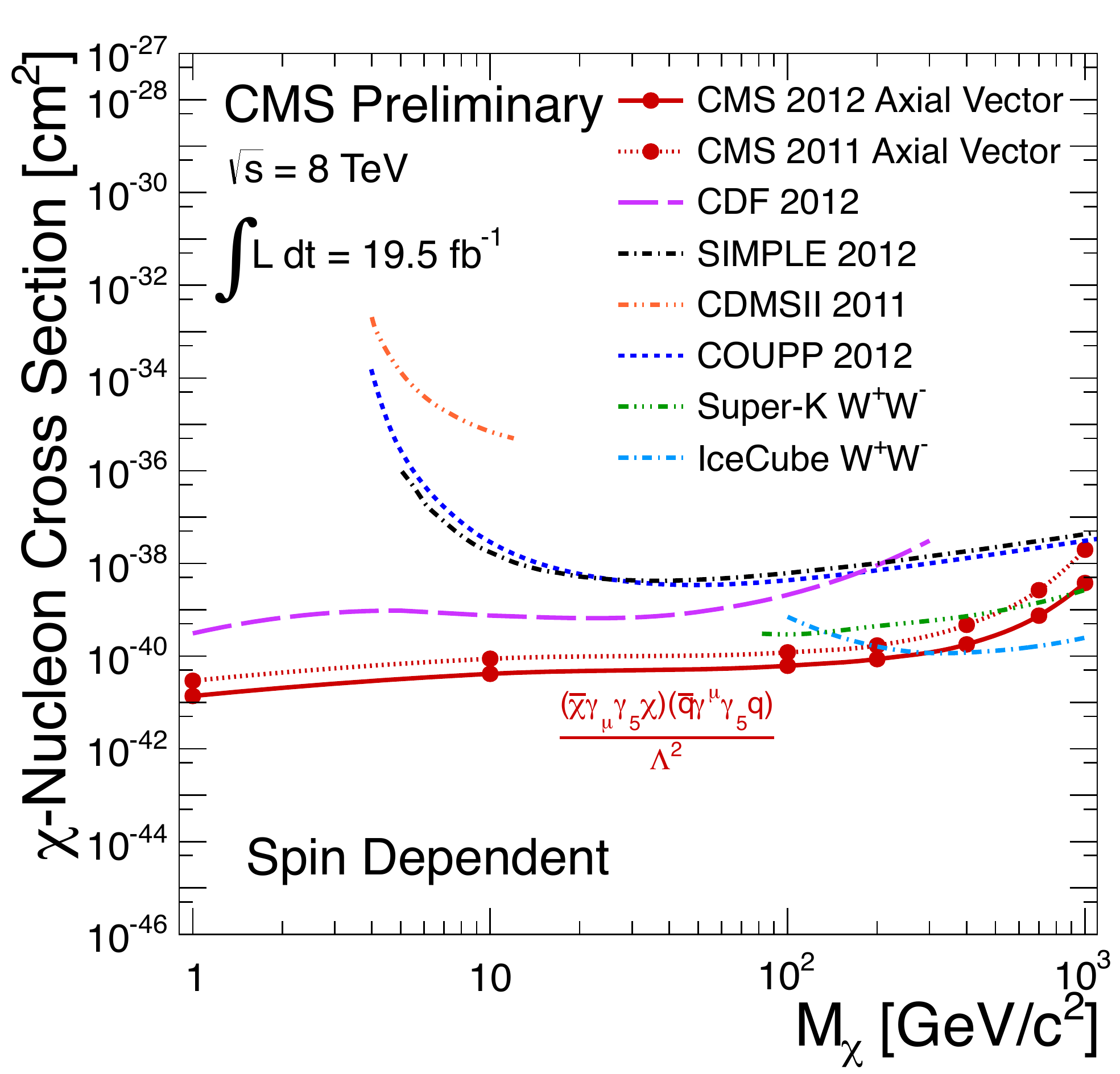}
    \vspace*{-0.4cm}
    \captionof{figure}{Constraints on DM production at the LHC from (left) ATLAS~\protect\cite{ATLAS-DM} and (right) CMS~\protect\cite{CMS-DM} monojet searches at $\sqrt{s} = 8$ TeV. For detailed description of the plots, see the corresponding references.}
\label{fig:DM}
\end{figure}

There are other proposed natural solutions to the hierarchy problem that do not invoke SUSY, e.g. extra dimensions in space, little Higgs models, etc. Searches for those so far also came empty-handed, but the situation is similar to the natural SUSY searches: a high-energy LHC run at 13--14 TeV is needed to fully explore the natural model space.

\section{Toward the Future}

What does it take to go to a higher energy at the LHC? The accelerator complex is currently being upgraded, and so are the LHC experiments. The main work being done on the LHC machine is the refurbishment of the interconnects between the dipole magnets in the machine, which proved to be insufficient to increase the energy of the machine beyond 8 TeV without a risk of a breakdown. This work is being conducted on over 10,000 superconducting splices.

The machine is expected to come back online in 2015 at an energy of $\sqrt{s} \sim 13$ TeV, close to the design energy of 14 TeV. Further increase in energy would require training of a number of dipole magnets via repeated quenches, and depending on the number of quenches needed the energy is expected to reach 13.5-14.0 TeV in subsequent years. The machine will be switched from the current 50 ns operation mode to a 25 ns beam crossing time which would allow to reach instantaneous luminosities $\sim 2 \times 10^{34}$ cm$^{-2}$s$^{-1}$ with the pileup of $\sim 50$ interactions per beam crossing. The current LHC schedule anticipates the delivery of 300--400\ifb to ATLAS and CMS by 2022.

At that point many elements of the machine will be damaged by radiation and will need to be replaced. In addition, the LHC instantaneous luminosity will saturate and the data doubling time will increase considerably. The detectors will also suffer significant radiation damage and will need to undergo a major upgrade. This situation makes a strong case for a High-Luminosity LHC (HL-LHC) upgrade around 2023. The goal of the HL-LHC is to reach the peak luminosity of $10^{35}$ cm$^{-2}$s$^{-1}$ with proposed running mode with the luminosity continuously leveled at half the peak value. That would allow the LHC to deliver $\sim 3000\ifb$ to the experiments by 2036.

Both ATLAS and CMS are planning major ``Phase 2" upgrades for the HL-LHC era. The goal of these upgrades is to replace components of the detector, which will reach the end of the life cycle due to radiation damage. That includes the entire central tracking systems and forward calorimetry. One also has to prepare the detectors for much harsher running conditions at the HL-LHC (up to a factor of five higher pileup), which requires redesigned trigger and DAQ systems, possible use of fast timing for pileup mitigation, hardware tracking trigger, and improved forward detectors needed for tagging vector boson fusion processes via forward jet detection. The goal is to achieve the same or better performance as in Run 1 under the HL-LHC conditions. The ALICE and LHCb experiments are considering upgrades for the HL-LHC running as well.

In order to build a strong physics case for the HL-LHC, the ATLAS and CMS experiments have conducted a number of studies on the reach of the HL-LHC in terms of searches for new physics and measurements of the properties of the Higgs boson. The results of these studies~\cite{ESPG} have been submitted to the European Strategy Planning Group (ESPG) in October 2012. The main physics goals of the HL-LHC are to determine the couplings of the Higgs boson to photons, gluons, $W/Z$ bosons, top- and bottom-quarks, and $\tau$-leptons to a few percent level that would allow to either see deviations from the SM Higgs properties, or to prove that the Higgs boson we discovered is indeed the SM Higgs boson. Other important goals of the HL-LHC is to measure the coupling of the Higgs boson to muons, as well as the Higgs self-coupling, and to observe unitarization of the vector boson scattering at high energy in the presence of the Higgs boson.

Higgs physics is not the only strong case for the HL-LHC. Other physics topics for which HL-LHC is required include finding massive new physics or ruling out a broad class of natural new physics models and demonstrating that SM is fine-tuned; answering the major question if we have entered the energy ``desert" and there are no new weakly or strongly interacting states below a few TeV; and probing higher energy scales via precision measurements. In case new physics will be found before the HL-LHC upgrade the extended running will be needed to measure the properties of new particles. All of these goals make the physics case for the HL-LHC very strong.

\section{Conclusions}

The LHC is the most successful and amazing particle accelerator built so far. The first three years of spectacular performance of the machine and the detectors brought in the first major discovery and a whole new program of precision measurements and searches. The LHC is taking a short break till 2015 to come back at $\sqrt{s} \sim 13$ TeV energy to explore the Terascale with a full potential. Running beyond 2022 with ten times higher integrated luminosity (HL-LHC) will be needed for detailed studies of the Higgs sector and any new physics to be found beforehand. The LHC is a very young machine, and it has a 20+ year long exciting program ahead, which is what we need to fully explore the properties and the consequences of the new particle we just discovered.

\section*{Acknowledgments}

I'd like to thank the organizers for a kind invitation and warm hospitality! I'd also like to thank my many colleagues in ALICE, ATLAS, CMS, LHCb, LHCf, and TOTEM experiments for producing beautiful results, only small fraction of which has been covered in this talk, for helpful discussions, and for providing some of the plots. This work has been partially supported by the DOE Grant \#DE-SC0010010.

\section*{Appendix}

Between May and September of 2013 a number of results described in these proceedings have undergone major updates. The main changes include the observation of the $B^0_S \to \mu\mu$ decay by the CMS and LHCb Collaborations~\cite{Bsmm-obs}, publication of the preliminary CMS result~\cite{CMS-polar} with an addition of the $J/\psi$ polarization measurement~\cite{CMS-polar-pub}, an indication of the $\Upsilon(nS)$ suppression in $p$Pb collisions~\cite{CMS-Upsilon-pPb}, publication of the ATLAS $H(\gamma\gamma)$, $H(ZZ)$, and $H(WW)$ searches~\cite{ATLAS-paper1} and the spin-parity determination~\cite{ATLAS-paper2}, as well as significantly expanded searches for new physics and dark matter~\cite{DM-new}. The ATLAS and CMS ESPG HL-LHC studies have been superseded by more detailed ones submitted as the Snowmass 2013 white papers~\cite{Snowmass}.

\end{document}
